\documentclass[apl,english]{revtex4-1}
\usepackage{amsmath, xcolor}
\usepackage{graphicx}
\usepackage{amssymb}
\usepackage{epsfig}

\usepackage{babel}
\usepackage{dsfont}

\renewcommand{\Im}{{\rm Im}}
\newcommand{\rd}{{\rm d}}
\newcommand{\kb}{k_{\rm B}}

\newcommand{\rs}{{\rm s}}
\newcommand{\rp}{{\rm p}}

\newcommand{\re}{{\rm e}}

\newcommand{\rr}{{\rm r}}
\newcommand{\rl}{{\rm l}}

\newcommand{\rth}{{\rm th}}

\begin{document}

\title{Coherent thermal conductance in multilayer photonic crystals}

\author{M. Tschikin}
\affiliation{Institut f\"{u}r Physik, Carl von Ossietzky Universit\"{a}t,
D-26111 Oldenburg, Germany.}

\author{P. Ben-Abdallah}
\affiliation{Laboratoire Charles Fabry, Institut d'Optique, CNRS, Universit\'{e} Paris-Sud, Campus
Polytechnique, RD128, 91127 Palaiseau Cedex, France}

\author{S.-A. Biehs}
\affiliation{Institut f\"{u}r Physik, Carl von Ossietzky Universit\"{a}t,
D-26111 Oldenburg, Germany.}

\date{\today}

%\pacs{44.40.+a, 78.20.-e, 78.67.-n, 03.50.De}
\begin{abstract}
We present an exact calculation of the coherent thermal conductance 
in a 1-D multilayer photonic crystals (PC) using the S-matrix method. 
In particular, we study the thermal conductance in a bilayer 
structure of slabs of Si/vacuum or Al$_2$O$_3$/vacuum by means of the exact expression for the radiative 
heat flux. We compare our results with results obtained in previous works. Our results show that the coupling of surface modes as well as material losses  play a fundamental role in the definition of the thermal conductance of PCs 
\end{abstract}

\maketitle

Recently there has been a growing interest in exploring nanoscale heat transfer 
theoretically and experimentally which is triggered by the fact that radiative 
heat flux at the nanoscale can be much larger than that between two black 
bodies~\cite{Polder1971} and quasi-monochromatic~\cite{SurfaceScienceReports,Volokitin2007} which makes it very 
promising for near-field thermophotovoltaics~ 
%\cite{MatteoEtAl2001,NarayanaswamyChen2003,LarocheEtAl06,ParkEtAl2007,ZhangReview}. 
\cite{ZhangReview}.
The tremendous increase in the amount of transfered energy for distances much smaller than
the thermal wavelength ($\lambda_\rth = \hbar c/\kb T$) which can be several orders
of magnitude larger than the value predicted by Stefan-Boltzmann's law 
can for dielectrics be attributed to the contribution of a large number of 
coupled surface phonon polariton modes~\cite{MuletAPL,BiehsPRL2010}.

The contribution of surface modes (SMs) is indeed very important for nanoscale heat fluxes and
many researchers have tried to enhance the amount of transfered heat by using this effect.
Volokitin and Persson~\cite{Volokitin2004} have pointed out that thin metallic
coatings on a substrate can increase the nanoscale heat flux, Biehs {\itshape et al.}~\cite{BiehsEtAl2007,Biehs2007} and 
Francoeur {\itshape et al.}~\cite{FrancoeurAPL} have shown that one can use the coupling of SMs in 
thin metallic or dielectric films to enhance the nanoscale heat flux, 
Ben-Abdallah and coworkers~\cite{PBA2009,PBA2010,Pryamikov2011,MessinaEtAl2012} and Francoeur {\itshape et al.}~\cite{Francoeur2009} have also
considered this effect between two finite slabs or media with several layers, 
Fu and Zhang~\cite{Zhang2005} have studied how doping affects the surface mode contribution,
van Zwol {\itshape et al.}~\cite{Zwol2010b} have shown that large nanoscale heat fluxes
in phase change materials are due to SMs, and Svetovoy {\itshape et al.}~\cite{Svetovoy2011} and
Ilic {\itshape et al.}~\cite{IlicEtAl2012} have pointed out that thin sheets of graphene allow do control or
modulate the surface mode contribution. Very recent works have also considered heat fluxes
for artificial structures and/or meta-materials supporting SMs in the infrared 
regime~\cite{Joulain2010,Zheng2011,FrancoeurEtAl2011,LussangeEtAl2012,BiehsEtAl2011}, or the surface mode coupling in 
many particle systems~\cite{PBAEtAl2011}. 

%It should also be mentioned that very recent experiments could unambiguously show
%that the radiative heat flux increases for distances smaller than the thermal wavelength~\cite{Kittel,HuEtAl2008,NanolettArvind},
%and can exceed the Stefan-Boltzmann law~\cite{ShenEtAl2008,Ottens2011}. The effect of modulation of heat fluxes
%by means of phase-change materials could also be demonstrated very recently~\cite{vanZwolEtAl}. The overall 
%agreement between the theoretical predictions based on the fluctuational electrodynamics~\cite{Polder1971} 
%and the experimental results is very good~\cite{NatureEmmanuel,Ottens2011,vanZwolEtAl}.

\begin{figure}[Hhbt]
  \epsfig{file=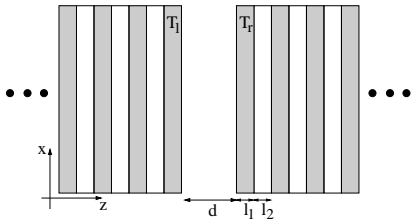, width= 0.3\textwidth}
  \caption{\label{Fig:SketchPhotonCryst} Sketch of the situation considered here. Two identical 
            1-D bilayer PCs are separated by a vacuum gap of distance $d$. Here we choose $d = l_2$
            to determine the thermal conductance inside an infinite 1-D PC.}
\end{figure}

In this letter, we will revisit the theory of thermal conductance by photons within a PC 
as depicted in Fig.~\ref{Fig:SketchPhotonCryst}. We will provide an
exact expression for the thermal conductance inside a 1-D PC for arbitrary 
dispersive and dissipative material slabs. In particular, this allows us to determine the transmission 
coefficients (TC) for the Bloch states inside the PC. In previous works Lau and 
co-workers~\cite{LauEtAl2008} have assumed that the TC equals its 
maximum value of one when losses can be neglected. Surprisingly, by comparing the exact results 
of our calculation with the results of Ref.~\cite{LauEtAl2008} we find that in the limit of 
vanishing losses the TC for the total internal reflection modes 
goes to zero and not to its maximum value. In fact, we find that the TC is 
very sensitive to the losses inside the PC slabs. In addition, our exact expression 
takes the contribution of SMs to the thermal conductance inside the PC into 
account as well. We will show that this surface mode contribution can be crucial for the 
thermal conductance inside a PC.

In order to derive the expression for the radiative heat flux inside a PC,
we assume that we have first two semi-infinite PCs as depicted in Fig.~\ref{Fig:SketchPhotonCryst}. The bilayer
structure has a period of $a = l_1 + l_2$ where $l_1$ is the thickness of the material
layer with complex permittivity $\epsilon_1=\epsilon_1'+{\rm i}\epsilon_1''$ and $l_2$ is the thickness of the vacuum layer
with $\epsilon_2 = 1$.  
The heat flux 
\begin{equation}
  \phi_{\rl\rr} = h_{\rl\rr}(T_l,d) \Delta T_{\rl\rr}
\end{equation}
between two such semi-infinite structures having a temperature difference $\Delta T_{\rl\rr} = T_\rr - T_\rl$ 
across the vacuum gap of distance $d$ can be derived from the expression given by Polder and van Hove~\cite{Polder1971}. 
The heat transfer coefficient (HTC) within an infinite PC can then be obtained by setting $d = l_2$ 
\begin{equation}
  h_{\rl\rr} = \sum_{i = \rs,\rp} \int_0^\infty\!\!\frac{\rd \omega}{2 \pi} \frac{\partial \Theta(T)}{\partial T}\biggr|_{T_\rl}
               \int\!\!\frac{\rd^2 \kappa}{(2 \pi)^2}  \mathcal{T}_i(\omega,\kappa;d = l_2).
\label{Eq:htc}
\end{equation}
Here, the time derivative of the Bose-Einstein function is given by
$\partial \Theta(T)/\partial T = (\hbar \omega)^2/(\kb T^2)\re^{\hbar \omega/\kb T}/(\re^{\hbar \omega/\kb T} - 1)^2$
%\begin{equation}
%   \frac{\partial \Theta(T)}{\partial T} = \frac{(\hbar \omega)^2}{\kb T^2} \frac{\re^{\hbar \omega/\kb T}}{(\re^{\hbar \omega/\kb T} - 1)^2}
%\end{equation}
and evaluated at the temperature $T_\rl$ of the last slab of the PC at the left hand side.
The TCs $\mathcal{T}_{i}(\omega,\kappa; d)$ for s- and p-polarized waves ($i = \rs, \rp$) are given by~\cite{Polder1971}
\begin{equation}
    \mathcal{T}_{i}(\omega,\kappa; d) =
    \begin{cases}
     \frac{(1 - |R^\rl_j|^2) (1 - |R^\rr_j|^2)}{|D^{\rl\rr}_j|^2}, & \kappa < \omega/c\\
     \frac{4 \Im(R^\rl_j)\Im(R^\rr_j){\rm e}^{-2 |k_{z0}| d}}{|D^{\rl\rr}_j|^2} ,  & \kappa > \omega/c
  \end{cases}
\label{Eq:transmcoeff}
\end{equation}
where $D^{\rl\rr}_i = (1 - R^\rl_i R^\rr_i {\rm e}^{2 {\rm i} k_{z0} d})^{-1}$
is a Fabry-P\'{e}rot-like denominator with $k_{z0}^2 = \omega^2/c^2 - \kappa^2$ and $\kappa^2 = k_x^2 + k_y^2$. 
$R_\rs$ and $R_\rp$ are the reflection coefficients for the two semi-infinite
PCs and can be calculated with the standard S-matrix method for layered 
media~\cite{Yeh,PBA2009,FrancoeurAPL}. Note, that the TC is for propagating modes with parallel wave vectors
$\kappa < \omega/c$ different from the expression for evanescent modes with parallel wave vectors $\kappa > \omega/c$.

Now, we are in a position to compare results from the exact expression in Eq.~(\ref{Eq:htc}) with the 
results in Ref.~\cite{LauEtAl2008}. First we note, that in the approach 
in Ref.~\cite{LauEtAl2008} the authors assume that
the TCs equal their maximum value of one for all propagating bloch modes inside the
PC. That means, the integral over all parallel wave vectors $\kappa$ is replaced by
\begin{equation}
  A(\omega) = \sum_{i = \rs,\rp} \int\!\!\frac{\rd^2 \kappa}{(2 \pi)^2}  \mathcal{T}_i(\omega,\kappa;d = l_2) 
            \rightarrow  2 \int'\!\!\frac{\rd^2 \kappa}{(2 \pi)^2} = A'(\omega).
\label{Eq:A_omega}
\end{equation}
Here the prime notes that the integral is for each frequency $\omega$ carried out
over the whole parallel wave vector range which allows for propagating solutions inside the PC, i.e.,
over the photonic Bloch bands. The photonic Bloch bands can be determined from the dispersion 
relation for the Bloch modes (see Ref.~\cite{Yeh}).
%\begin{equation}
%\begin{split}
%  \cos(k_z \Lambda) &= - \frac{1}{2}\biggl(\frac{k_{z1}}{P_i k_{z2}} + \frac{P_i k_{z2}}{k_{z1}} \biggr) \sin(k_{z1} l_1) \sin(k_{z2} l_2)\\
%                    &\quad + \cos(k_{z1} l_1) \cos(k_{z2} l_2).
%\end{split}
%\label{Eq:BlochDisp}
%\end{equation}
%Here, $k_{z1}^2 = \epsilon_1 \omega^2/c^2 - \kappa^2 = $, $k_{z2}^2 = k_{z0}$, 
%and $P_\rs = 1$ and $P_\rp = \epsilon_1$.
Hence, the results from Refs.~\cite{LauEtAl2008} for the coherent thermal conductance give the upper limit
for the contribution of the propagating Bloch modes. As we will see in the following, the exact result can be
very different from such a calculation due to losses, resonant SMs as well as evanescent Bloch modes.   

\begin{figure}[Hhbt]
  \epsfig{file=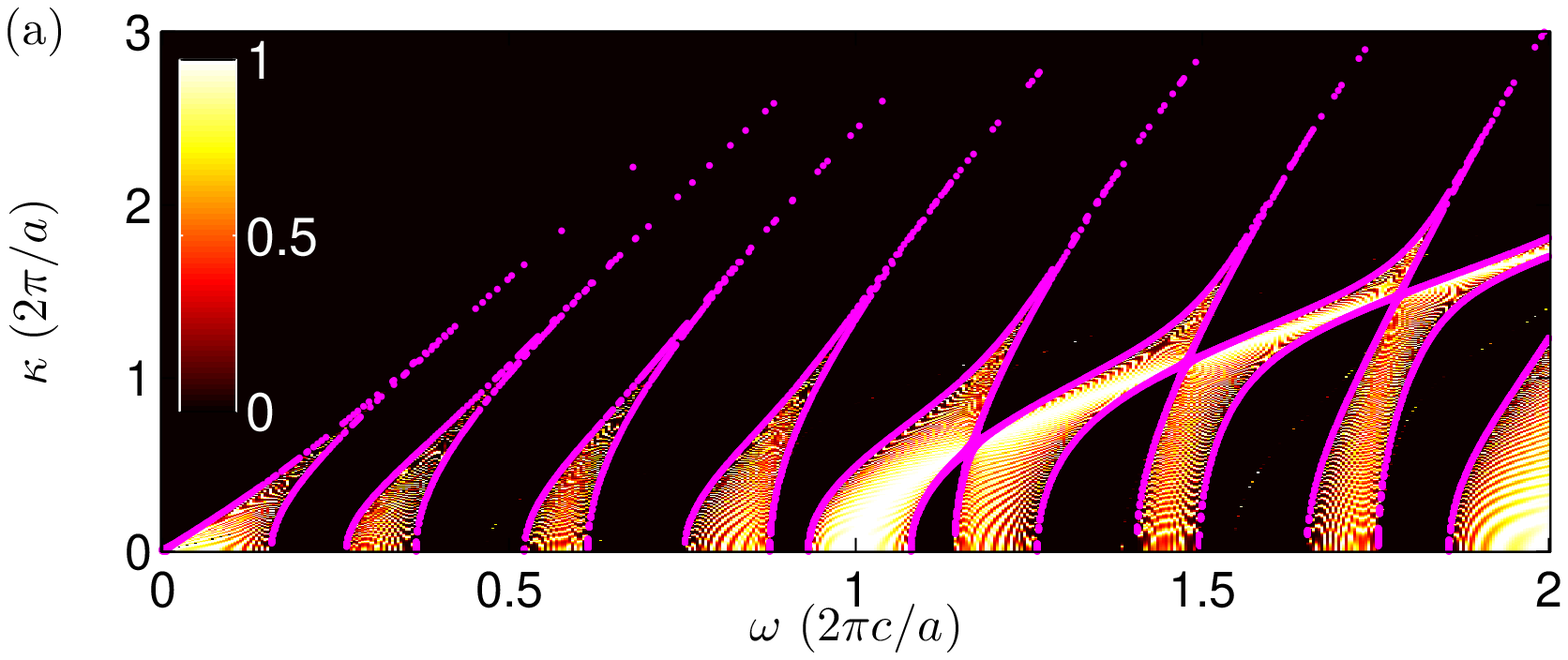, width= 0.4\textwidth}
  \epsfig{file=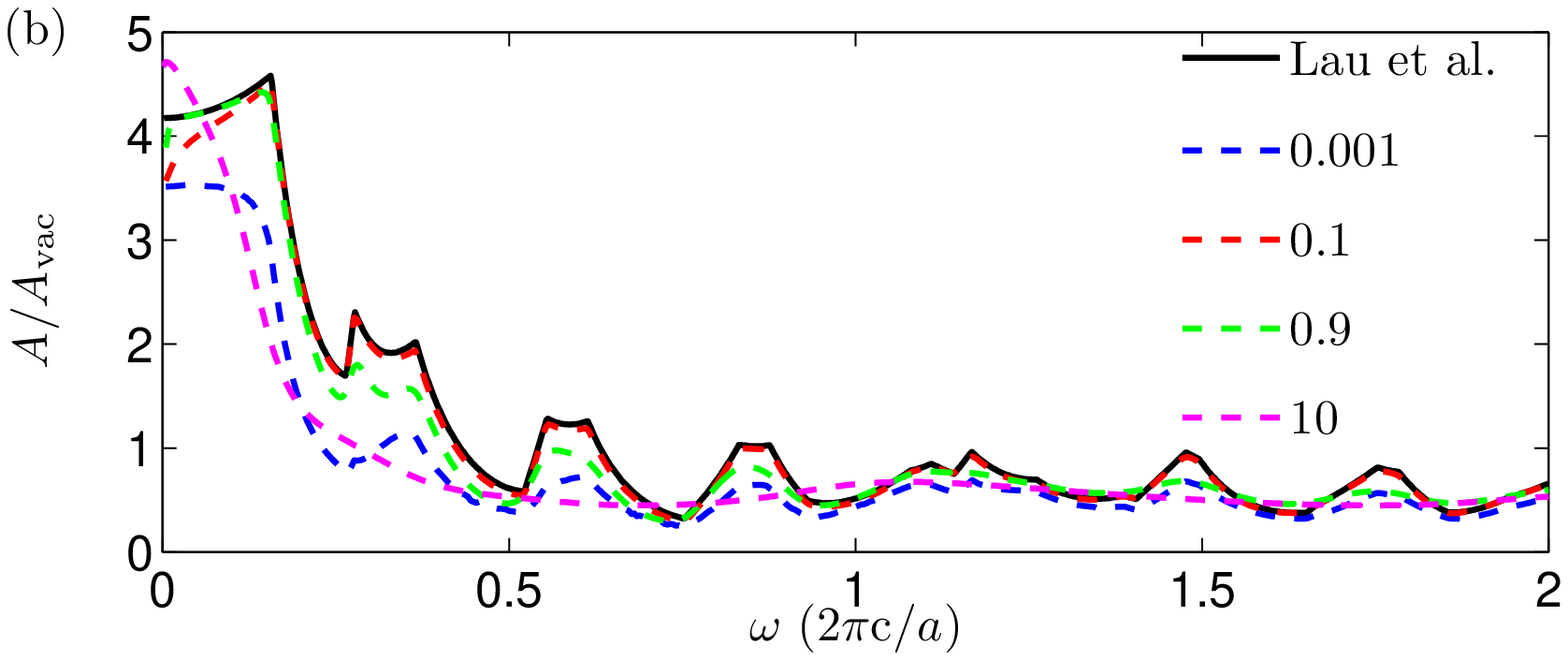, width= 0.4\textwidth}
  \epsfig{file=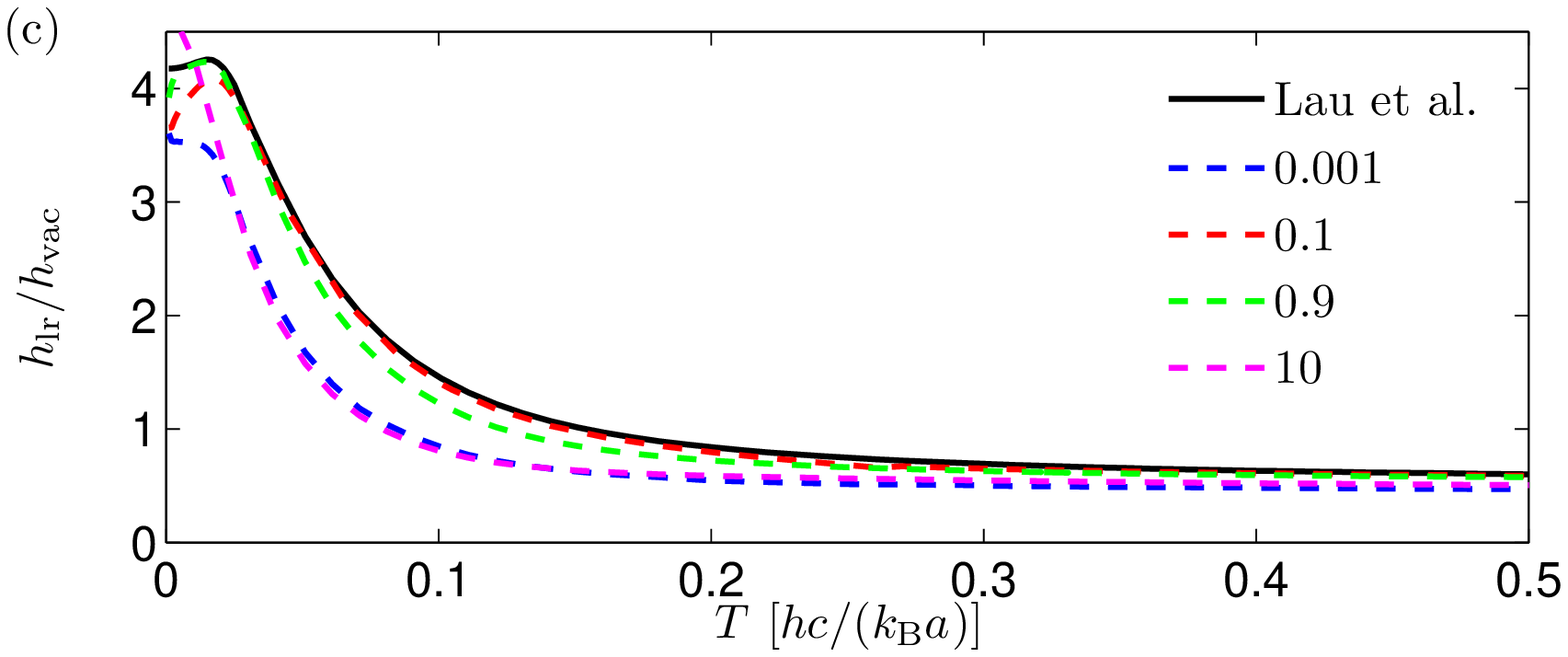, width= 0.4\textwidth}
  \caption{\label{Fig:TrCo_Aomega_hTemp} (Color online) (a) The TC for s-polarized modes defined in Eq.~(\ref{Eq:transmcoeff}) in $\omega$-$\kappa$ plane for \mbox{$\epsilon_1=12+0.001\cdot{\rm i}$}. In magenta: numerical results for the boundaries of the Bloch mode dispersion relation~\cite{Yeh}. (b) $A(\omega)$ from Eq. (\ref{Eq:A_omega}) normalized to $A_{\rm vac}=(\omega/c)^2 /(2\pi)^2$. The solid line represents the result for $\epsilon_1'= 12$ and $\mathcal{T}_{\rm s}=\mathcal{T}_{\rm p}=1$ for Bloch modes. The dashed lines show the exact results for fixed $\epsilon_1'= 12$ and different $\epsilon_1''$. (c) The HTC for the same permittivities as in (b) normalized to the black body result.}
\end{figure}

In Fig.~\ref{Fig:TrCo_Aomega_hTemp}(a) the trans\-mission coefficient $\mathcal{T}_\rs$ is plotted in the $\omega$-$\kappa$ plane
choosing $\epsilon_1=12+{\rm i}\cdot0.001$ for a PC with $100$ slabs. 
It can be seen that although the imaginary part of the permittivity is very 
small, corresponding to a system with vanishing losses, the TC is less than one for most parts of the Bloch bands.
We find similar results for the p-polarized modes. In Fig.~\ref{Fig:TrCo_Aomega_hTemp}(b) we present the numerical results
for $A(\omega)$ when integrating the TC over $\kappa$ using Eq. (\ref{Eq:A_omega}). The plotted values
are normalized to the maximum value possible for propagating modes $A_{\rm vac}=\frac{1}{2\pi}(\omega/c)^2 $ inside the vacuum gap.
The solid black line represents the result from Ref.~\cite{LauEtAl2008} and the colored dashed curves represent the exact 
results using the same $\epsilon_1'$ as in Ref.~\cite{LauEtAl2008} but for different $\epsilon_1''$. The best agreement with 
the black curve is found for $\epsilon_1''=0.1$. When decreasing the losses by making $\epsilon_1''$ smaller than $\epsilon_1'' = 0.1$ then
$A(\omega)$ decreases as well for nearly all frequencies so the deviation from the black curve gets larger. 
This means that for vanishing losses the TC does not converge to its maximum value for all Bloch modes. 
On the other hand, when making $\epsilon_1''$ larger than $\epsilon_1'' = 0.1$ the skin depth $\delta_{\rm s}= 1/\bigl[\omega/c {\rm Im}(\sqrt{\epsilon_1})\bigr]$ inside the material slabs 
decreases and attains for $\epsilon_1''=10$ (dashed magenta curve) values on the order of the 
period $a$ of the PC so that the field is damped at this scale. Hence, the heat flux is not coherent anymore and the Bloch band structure in $\mathcal{T}_{\rm s/p}$
disappears. In fact, then the heat flux is due to Fabry-P\'{e}rot modes of the cavity formed by separation gap and explaine the smooth and weak oscillating behaviour of $A(\omega)$ when $\epsilon_1''=10$ [Fig.~\ref{Fig:TrCo_Aomega_hTemp}(b)]. Finally, in Fig.~\ref{Fig:TrCo_Aomega_hTemp}(c) we show the HTC 
$h_{\rm lr}$ versus the temperature $T$ for the same permittivities as in Fig.~\ref{Fig:TrCo_Aomega_hTemp}(b) 
nor\-ma\-li\-zed to the vacuum or black body value \mbox{$h_{\rm vac}=(\pi^2 k_{\rm B}^4T^3)/(15c^2\hbar^3)$}. 
Especially, for small $T$ the deviation with respect to the results of Ref.~\cite{LauEtAl2008} is relatively large, whereas
for large temperatures we obtain values very close to the 'universal' value found in Ref.~\cite{LauEtAl2008}. 

\begin{figure}[Hhbt]
  \epsfig{file=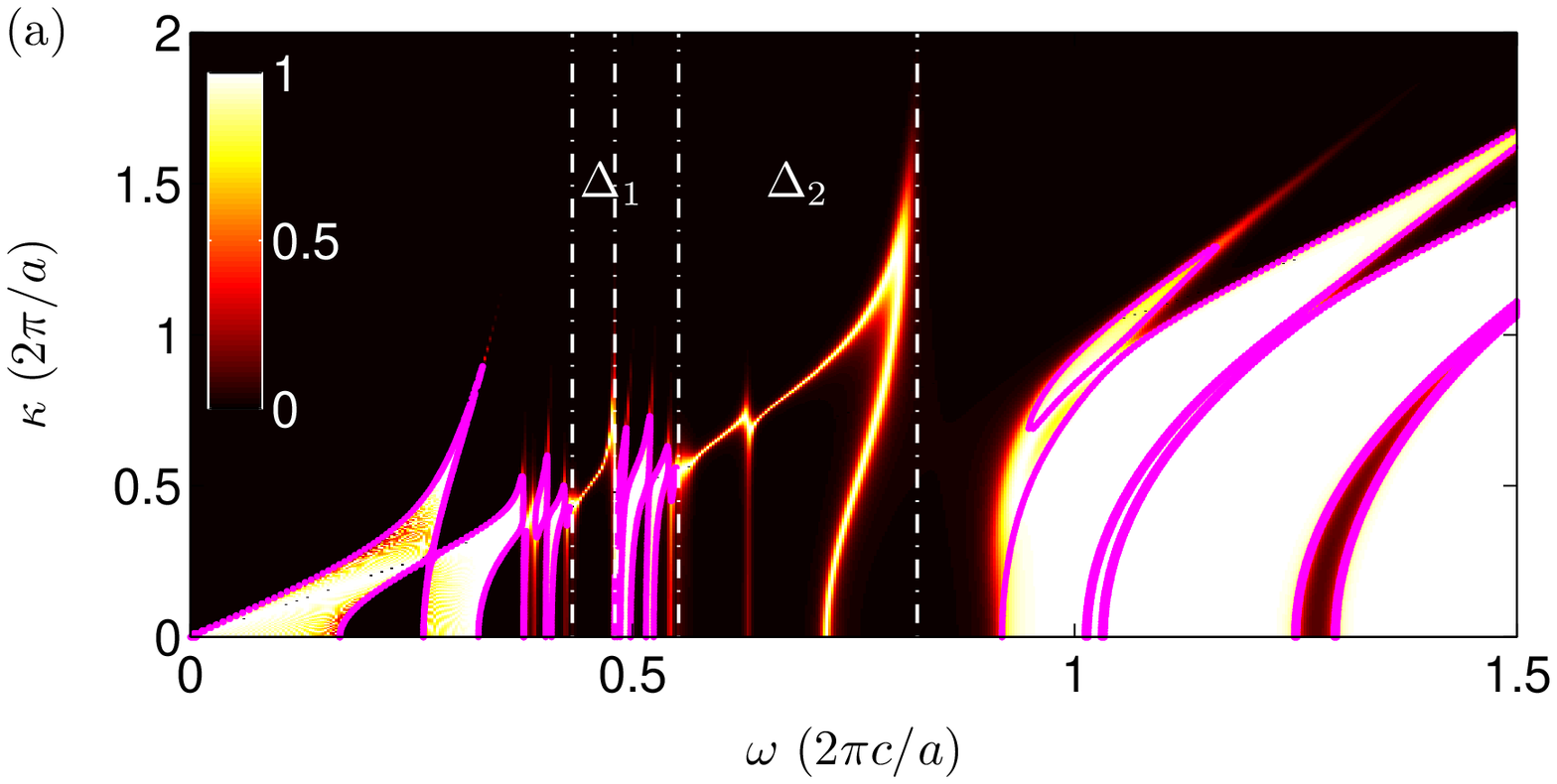, width= 0.4\textwidth}
  \epsfig{file=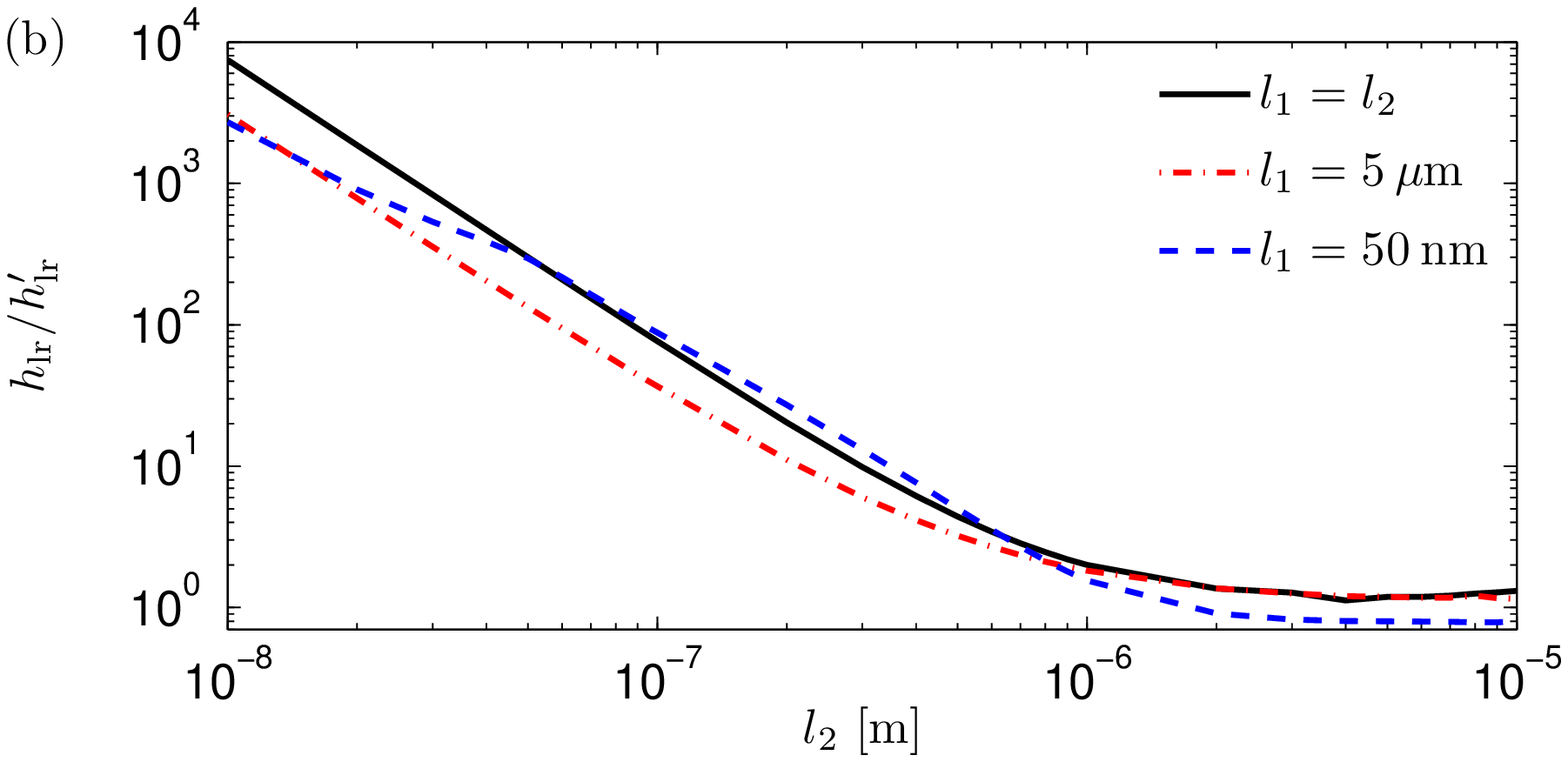, width= 0.4\textwidth}
  \caption{\label{Fig:Al2o3} (Color online) (a) The TC for p-polarized modes defined in Eq.~(\ref{Eq:transmcoeff}) in $\omega$-$\kappa$ plane for Al$_2$O$_3$/vacuum structure with $l_1=l_2=5\,\mu$m. In magenta: numerical results for the boundaries of the Bloch mode dispersion relation~\cite{Yeh}. The white dashed lines mark the frequency regions $\Delta_1$ and $\Delta_2$ where coupled SPPs exist. (b) HTC $h_{\rm lr}$ normalized to $h_{\rm lr}'$ where $A(\omega)=A'(\omega)$ and plotted vs. the vacuum gap $l_2$. For the solid curve the Al$_2$O$_3$ slab $l_1$ varies analog to $l_2$ and the dashed curves show the results for a fixed $l_1 = 5\,\mu{\rm m}$ and $l_1 = 50\,{\rm nm}$ while varying $l_2$.}
\end{figure}

Here below we examine the behaviour of structures which are able to support SMs, surface phonon polaritons (SPPs). To do that we consider an Al$_2$O$_3$/vacuum PC with 100 slabs at $T_l=300\,$K. For this material combination SPPs not only exist for the p-polarized modes but also play the important role for heat transfer at subwavelength distances. In Fig.~\ref{Fig:Al2o3}(a) we have plotted the TC for p-polarized modes in $\omega$-$\kappa$ plane. It is obvious that not only the Bloch modes and Bloch SMs contribute to the heat conductance but also coupled SPP modes which can be identified in the frequency bands $\Delta_1$ and $\Delta_2$ where $\epsilon_1 < -1$.  To compare our exact calculations with results from~\cite{LauEtAl2008} for the Al$_2$O$_3$/vacuum PC we have plotted in Fig.~\ref{Fig:Al2o3}(b) the HTC $h_{\rm lr}$ versus the vacuum gap $l_2$ for different $l_1$. The results are normalized to the HTC $h_{\rm lr}'$ from Ref.~\cite{LauEtAl2008} for which $A(\omega)=A'(\omega)$ [see Eq.~(\ref{Eq:A_omega})]. It can be seen that the exact HTC can be nearly four orders of magnitude larger than the HTC calculated with the approximative method at $l_2=10\,$nm. This can be attributed to the SPP mode contribution which is proportional to  $1/l_2^2$ for small gap sizes.

%\begin{acknowledgements}
  M.\ T.\ gratefully acknowledges support from the Stif\-tung der Metallindustrie im Nord-Westen. P. B. A. acknowledges the support of the Agence Nationale de la Recherche through the Source-TPV project ANR 2010 BLANC 0928 01.
%\end{acknowledgements}

\end{document}